\documentclass[preprint,showpacs,preprintnumbers,amssymb,aps]{revtex4}
\usepackage{amsfonts,color,graphicx,epsfig,amsmath,multirow,slashed}
\usepackage[colorlinks=true,linkcolor=blue,citecolor=blue, urlcolor=blue]{hyperref}

\begin{document}

\title{Further studies on the exclusive productions of $J/\psi+\chi_{cJ}$ ($J=0,1,2$) via $e^+e^-$ annihilation at the $B$ factories}

\author{YingZhao Jiang}
\author{Zhan Sun}
\email{zhansun@cqu.edu.cn}

\affiliation{
\footnotesize
Department of Physics, Guizhou Minzu University, Guiyang 550025, P. R. China. }

\date{\today}

\begin{abstract}
By including the interference effect between the QCD and the QED diagrams, we carry out a complete analysis on the exclusive productions of $e^+e^- \to J/\psi+\chi_{cJ}$ ($J=0,1,2$) at the $B$ factories with $\sqrt{s}=10.6$ GeV at the next-to-leading-order (NLO) level in $\alpha_s$, within the nonrelativistic QCD framework. It is found that the $\mathcal O (\alpha^3\alpha_s)$-order terms that represent the tree-level interference are comparable with the usual NLO QCD corrections, especially for the $\chi_{c1}$ and $\chi_{c2}$ cases. To explore the effect of the higher-order terms, namely $\mathcal O (\alpha^3\alpha_s^2)$, we perform the QCD corrections to these $\mathcal O (\alpha^3\alpha_s)$-order terms for the first time, which are found to be able to significantly influence the $\mathcal O (\alpha^3\alpha_s)$-order results. In particular, in the case of $\chi_{c1}$ and $\chi_{c2}$, the newly calculated $\mathcal O (\alpha^3\alpha_s^2)$-order terms can to a large extent counteract the $\mathcal O (\alpha^3\alpha_s)$ contributions, evidently indicating the indispensability of the corrections. In addition, we find that, as the collision energy rises, the percentage of the interference effect in the total cross section will increase rapidly, especially for the $\chi_{c1}$ case.

\end{abstract}

\maketitle
\section{Introduction}

The exclusive production of double charmonia via the $e^{+}e^{-}$ annihilation at the $B$ factories is an ideal laboratory for the study of heavy quarkonium. In the first place, the process is ``clean". To be specific, the color-octet effect is negligible and the contributions of the color-singlet channels are dominant, which is beneficial to draw a definite conclusion. On the experiment side, the measurements on the total cross sections of $\sigma[e^+e^-\to J/\psi+\eta_c]$ and $\sigma[e^+e^-\to J/\psi+\chi_{c0}]$ \cite{Abe:2002rb,Abe:2004ww,Aubert:2005tj,Uglov:2004xa} both significantly overshoot the leading-order (LO) QCD predictions \cite{Braaten:2002fi,Liu:2002wq,Liu:2004ga,Hagiwara:2003cw} based on the nonrelativistic QCD framework \cite{Bodwin:1994jh}. In order to deal with the large discrepancy between theory and data, a great amount of attempts have been tried \cite{Prediction1,Prediction2,Prediction3,Prediction4,Prediction5,Prediction6,Prediction7,Zhang:2008gp}. Among them, the next-to-leading-order (NLO) QCD correction \cite{Prediction6,Prediction7,Zhang:2008gp} to the process is regarded as a breakthrough, significantly alleviating the tension between the theoretical predictions and the measured cross sections.

As pointed out in \cite{Braaten:2002fi,Liu:2004ga}, for $e^+e^-\to J/\psi+\eta_c$, in addition to the mentioned above essential NLO QCD corrections, the interference between the QCD and QED tree-level diagrams, namely the $\mathcal O (\alpha^3\alpha_s)$-order terms, can also provide significant contributions, which can be ascribed to the large kinematic enhancements caused by the single-photon-fragmentation (SPF) topologies of the QED diagrams. Moreover, recently Sun et al. \cite{Sun:2018rgx} find that the NLO QCD corrections to these $\mathcal O (\alpha^3\alpha_s)$-order terms can significantly further strengthen the effect of the interference terms.

Considering that the SPF topologies also exist in the process of $e^+e^-\to J/\psi+\chi_{c}$, the cross terms between the QCD and QED diagrams probably can as well have a significant effect on the total cross section, deserving a separate investigation. For this purpose, by introducing the interference terms up to the $\mathcal O (\alpha^3\alpha_s^2)$ order, we will carry out a further study on the exclusive production of $J/\psi+\chi_{c}$ via $e^+e^-$ annihilation at the $B$ factories, providing a complete comparison between the interference effects and the usual QCD contributions, at the QCD NLO level, for the first time.

The rest paragraphs are organized as follows: In Sec. II we give a description on the calculation formalism. In Sec. III, the phenomenological results and discussions are presented. Sec. IV is reserved as a summary.
\section{Calculation Formalism}

\begin{figure*}
\includegraphics[width=0.9\textwidth]{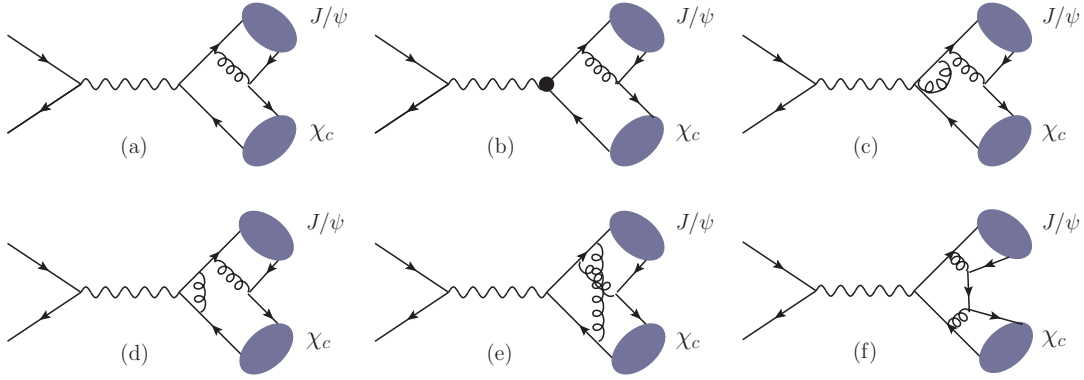}
\caption{Some sample QCD Feynman Diagrams for $e^+e^- \to J/\psi+\chi_c$. Fig.(\ref{fig:Feyn1}a) ($\mathcal M_{\alpha\alpha_s}$) is the QCD tree-level diagram. Figs.(\ref{fig:Feyn1}b-\ref{fig:Feyn1}f) ($\mathcal M_{\alpha\alpha_s^2}$) are NLO QCD corrections to Fig.(\ref{fig:Feyn1}a).}
\label{fig:Feyn1}
\end{figure*}

\begin{figure*}
\includegraphics[width=0.9\textwidth]{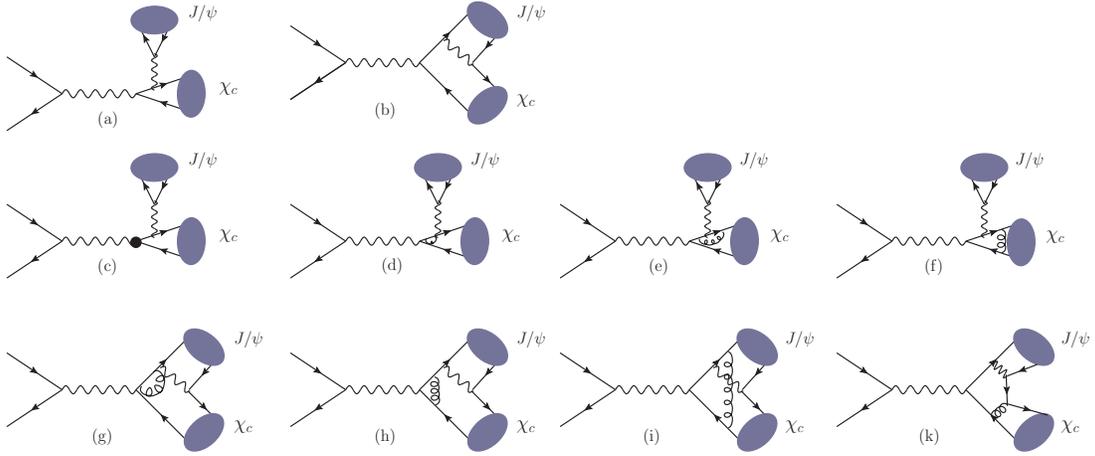}
\caption{Some sample QED Feynman Diagrams for $e^+e^- \to J/\psi+\chi_c$. Figs.(\ref{fig:Feyn2}a-\ref{fig:Feyn2}b) ($\mathcal M_{\alpha^2}$) are the QED tree-level diagrams, in which Fig.(\ref{fig:Feyn2}a) denotes the typical SPF diagram. Figs.(\ref{fig:Feyn2}c-\ref{fig:Feyn2}k) ($\mathcal M_{\alpha^2\alpha_s}$) are NLO QCD corrections to Fig.(\ref{fig:Feyn2}a) and Fig.(\ref{fig:Feyn2}b).}
\label{fig:Feyn2}
\end{figure*}

Up to the $\mathcal O (\alpha^3)$-order level, the squared matrix element of $e^{+}e^{-} \to J/\psi+\chi_c$ can be written as,
\begin{eqnarray}
&& |(\mathcal M_{\alpha\alpha_s}+\mathcal M_{\alpha\alpha_s^2})+(\mathcal M_{\alpha^2}+\mathcal M_{\alpha^2\alpha_s})|^2 \nonumber \\
 &=&|\mathcal M_{\alpha\alpha_s}|^2+2\textrm{Re}(\mathcal M_{\alpha\alpha_s}\mathcal M_{\alpha\alpha_s^2}^{*}) +2\textrm{Re}(\mathcal M_{\alpha\alpha_s}\mathcal M_{\alpha^2}^{*}) \nonumber \\
   &&+2\textrm{Re}(\mathcal M_{\alpha\alpha_s}\mathcal M_{\alpha^2\alpha_s}^{*}) +2\textrm{Re}(\mathcal M_{\alpha\alpha_s^2}\mathcal M_{\alpha^2}^{*})+\cdots
\end{eqnarray}
There are in total 84 QCD diagrams (4 tree-level, 60 one-loop and 20 counter-terms) and 72 QED diagrams (6 tree-level, 42 one-loop and 24 counter-terms) for $e^+e^- \to J/\psi+\chi_c$. Some sample Feynman diagrams are illustrated in Figs.(\ref{fig:Feyn1}) and (\ref{fig:Feyn2}). In calculating $\mathcal M_{\alpha^2\alpha_s}$, we will not need to carry out the NLO QED corrections to $\mathcal M_{\alpha\alpha_s}$, namely Fig.(\ref{fig:Feyn1}a), since these topologies are compensated by the initial-state radiation diagrams, which are irrelevant to the exclusive productions of $e^+e^- \to J/\psi+\chi_c$.

Ignoring the higher-order terms in $\alpha$, we divide the differential cross section into the following four parts:
\begin{equation}
d\sigma = d\sigma_{2}^{(0)}+d\sigma_{2}^{(1)}+d\sigma_{3}^{(0)}+d\sigma_{3}^{(1)}
\end{equation}
with
\begin{eqnarray}
d\sigma_{2}^{(0)} &\propto& |\mathcal M_{\alpha\alpha_s}|^2, \\
d\sigma_{2}^{(1)} &\propto& 2\textrm{Re}(\mathcal M_{\alpha\alpha_s}\mathcal M_{\alpha\alpha_s^2}^{*}), \\
d\sigma_{3}^{(0)} &\propto& 2\textrm{Re}(\mathcal M_{\alpha\alpha_s}\mathcal M_{\alpha^2}^{*}),  \\
d\sigma_{3}^{(1)} &\propto& 2\textrm{Re}(\mathcal M_{\alpha\alpha_s}\mathcal M_{\alpha^2\alpha_s}^{*})+2\textrm{Re}(\mathcal M_{\alpha^2}\mathcal M_{\alpha\alpha_s^2}^{*}).
\end{eqnarray}
The first two terms $d\sigma_{2}^{(0,1)}$ and the second two terms $d\sigma_{3}^{(0,1)}$ are the usual QCD contributions and the newly introduced interference terms up to NLO level in $\alpha_s$, respectively.

For the purpose of isolating the ultraviolet (UV) and infrared (IR) divergences, we will adopt the usual dimensional regularization procedure with $D=4-2\epsilon$. The on-mass-shell (OS) scheme is employed to set the renormalization constants of the charm-quark mass $Z_m$ and the filed $Z_2$, and the $\overline{\rm MS}$-scheme for the QCD gauge coupling $Z_g$ and the gluon field $Z_3$,
\begin{eqnarray}
\delta Z_{m}^{\rm OS}&=& -3 C_{F} \frac{\alpha_s N_{\epsilon}}{4\pi}\left[\frac{1}{\epsilon_{\textrm{UV}}}-\gamma_{E}+\textrm{ln}\frac{4 \pi \mu_r^2}{m_c^2}+\frac{4}{3}+\mathcal O(\epsilon)\right], \nonumber \\
\delta Z_{2}^{\rm OS}&=& - C_{F} \frac{\alpha_s N_{\epsilon}}{4\pi}\left[\frac{1}{\epsilon_{\textrm{UV}}}+\frac{2}{\epsilon_{\textrm{IR}}}-3 \gamma_{E}+3 \textrm{ln}\frac{4 \pi \mu_r^2}{m_c^2} \right. \nonumber\\
&& \left.+4+\mathcal O(\epsilon)\right], \nonumber \\
\delta Z_{3}^{\overline{\rm MS}}&=& \frac{\alpha_s N_{\epsilon}}{4\pi}(\beta_{0}-2 C_{A})\left[\frac{1}{\epsilon_{\textrm{UV}}}-\gamma_{E}+\textrm{ln}(4\pi)+\mathcal O(\epsilon)\right], \nonumber \\
\delta Z_{g}^{\overline{\rm MS}}&=& -\frac{\beta_{0}}{2}\frac{\alpha_s N_{\epsilon}}{4\pi}\left[\frac{1} {\epsilon_{\textrm{UV}}}-\gamma_{E}+\textrm{ln}(4\pi)+\mathcal O(\epsilon)\right],
\end{eqnarray}
where $\gamma_E$ is the Euler's constant, $\beta_{0}=\frac{11}{3}C_A-\frac{4}{3}T_Fn_f$ is the one-loop coefficient of the $\beta$-function and $n_f$ is the active quark flavor numbers, $N_{\epsilon}= \Gamma[1-\epsilon] /({4\pi\mu_r^2}/{(4m_c^2)})^{\epsilon}$. In ${\rm SU}(3)_c$, the color factors are given by $T_F=\frac{1}{2}$, $C_F=\frac{4}{3}$ and $C_A=3$.

\section{Phenomenological results and discussions}

Before presenting the phenomenological results, we first demonstrate the choices of the parameters in our calculations. The $e^+e^-$ collision energy is assumed to be $\sqrt{s}=10.6$ GeV. To keep the gauge invariance, both the masses of $J/\psi$ and $\chi_c$ are set to be $2m_c$, with $m_c=1.5$ GeV. $\alpha=1/137$. For the NLO calculations, we employ the two-loop $\alpha_s$ running, and one-loop $\alpha_s$ running for LO. The values of $|R_s(0)|^2$ and $|R^{'}_p(0)|^2$ are taken as $|R_s(0)|^2=0.81~\textrm{GeV}^3$ and $|R^{'}_p(0)|^2=0.075~\textrm{GeV}^5$, respectively \cite{Eichten:1995ch}.

As a cross check for our calculations, with the same choices of the input parameters, we have obtained the same NLO QCD predictions, namely $\sigma^{(1)}_{2}$, as those of Refs. \cite{Zhang:2008gp,Dong:2011fb}.

The total cross sections for $e^+e^- \to J/\psi+\chi_{cJ}$ ($J=0,1,2$) are presented in Table \ref{table:cross section}. One can see that, for the production of $J/\psi+\chi_{c0}$, the contributions of the $\mathcal O (\alpha^3\alpha_s)$-order terms representing the interference effect between the QCD and QED tree-level diagrams, namely $\sigma^{(0)}_3$, is about $6\%$ of the well-known positive and large NLO QCD corrections, $\sigma^{(1)}_2$. By calculating the NLO QCD corrections to $\sigma^{(0)}_3$, it is found that the newly obtained higher order terms, $\sigma^{(1)}_3$, can enhance $\sigma^{(0)}_3$ by about $4\% - 31\%$. In the case of $\chi_{c1}$, when $\mu_r=3$ GeV, the $\sigma^{(0)}_3$ is important, which is almost identical to $\sigma^{(1)}_2$, further reducing the LO QCD cross section, $\sigma^{(0)}_2$. To our astonishment, the newly caculated $\sigma^{(1)}_3$ can even reach up to $-113\%$ of $\sigma^{(0)}_3$, greatly compensating for the ``reduction" effect caused by $\sigma^{(0)}_3$. With regard to the production of $J/\psi$ in association with $\chi_{c2}$, $\sigma^{(0)}_3$ can also provide a sizeable contribution comparing to the usual NLO QCD corrections $\sigma^{(1)}_2$. And, similar to the $\chi_{c1}$ case, this significant $\sigma^{(0)}_3$ contribution will be still counteracted by $\sigma^{(1)}_3$ to a large extent. Therefore, to achieve a more precise prediction on the total cross sections for $e^+e^- \to J/\psi+\chi_{cJ}$, it is definitely indispensable to incorporate the new $\sigma^{(1)}_3$ ingredient by calculating the NLO QCD corrections to $\sigma^{(0)}_3$.
\begin{table*}[htb]
\centering
\caption{Total cross sections for $e^{+}e^{-} \to J/\psi+\chi_{cJ}$ (in unit: fb), with $J=0,1,2$. $m_c$=1.5 GeV. $\sqrt{s}=10.6$ GeV.}
\label{table:cross section}
\begin{tabular}{ccccccccccc}
\hline\hline
$J/\psi+\chi_{cJ}$ & $~~~~~\mu_r~~~$ & $~~\sigma_{2}^{(0)}$ & $~~~~\sigma_{2}^{(1)}~$ & $~~~~\sigma_{3}^{(0)}~$ & $~~~~\sigma_{3}^{(1)}~$ & $~\left|\frac{\sigma_{3}^{(0)}}{\sigma_{2}^{(1)}}\right|$ & $~\left|\frac{\sigma_{2}^{(1)}}{\sigma_{2}^{(0)}}\right|$ & $~\left|\frac{\sigma_{3}^{(1)}}{\sigma_{3}^{(0)}}\right|$ \\ \hline
$J=0$ & $2m_c$ & $6.62$ & $~~3.79$ & $~~0.24$ & $-0.01$ & $0.06$ & $0.57$ & $0.04$\\
$~$ & $\sqrt{s}/2$ & $4.48$ & $~~3.54$ & $~~0.19$ & $~~0.03$ & $0.05$ & $0.79$ & $0.16$\\
$~$ & $\sqrt{s}$ & $3.14$ & $~~3.11$ & $~~0.16$ & $~~0.05$ & $0.05$ & $0.99$ & $0.31$\\ \hline
$J=1$ & $2m_c$ & $1.12$ & $-0.10$ & $-0.08$ & $~~0.09$ & $0.80$ & $0.09$ & $1.13$\\
$~$ & $\sqrt{s}/2$ & $0.75$ & $~~0.19$ & $-0.06$ & $~~0.05$ & $0.32$ & $0.25$ & $0.83$\\
$~$ & $\sqrt{s}$ & $0.53$ & $~~0.28$ & $-0.05$ & $~~0.03$ & $0.18$ & $0.53$ & $0.60$\\ \hline
$J=2$ & $2m_c$ & $1.61$ & $-0.35$ & $~~0.13$ & $-0.10$ & $0.37$ & $0.22$ & $0.77$\\
$~$ & $\sqrt{s}/2$ & $1.09$ & $~~0.16$ & $~~0.11$ & $-0.05$ & $0.69$ & $0.15$ & $0.45$\\
$~$ & $\sqrt{s}$ & $0.76$ & $~~0.34$ & $~~0.09$ & $-0.02$ & $0.26$ & $0.45$ & $0.22$\\ \hline\hline
\end{tabular}
\end{table*}

\begin{figure*}[htb]
\includegraphics[width=0.59\textwidth]{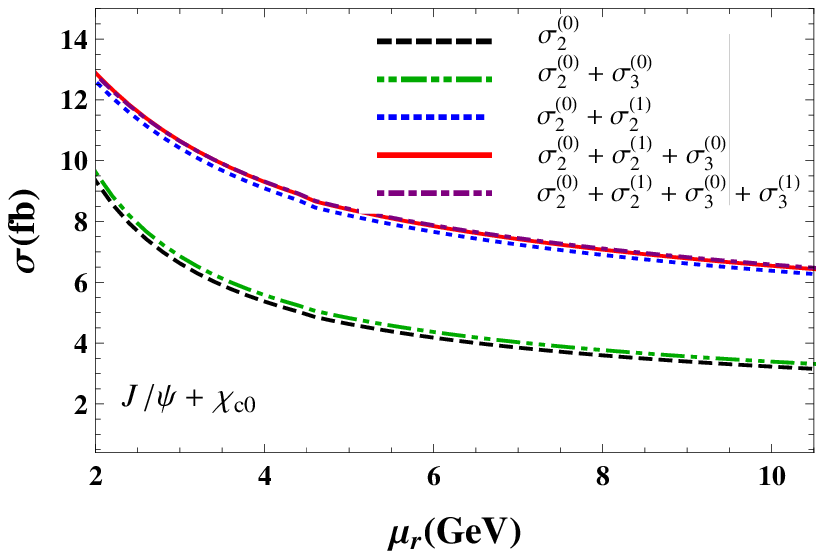}
\includegraphics[width=0.59\textwidth]{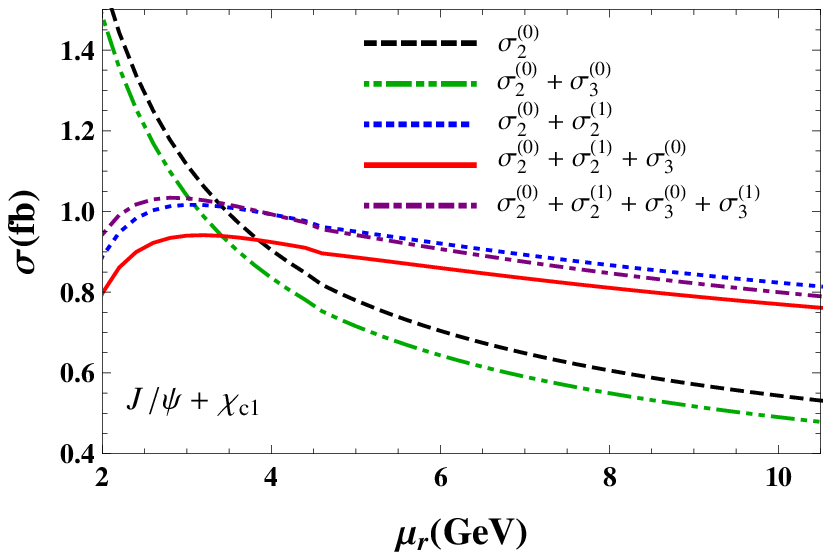}
\includegraphics[width=0.59\textwidth]{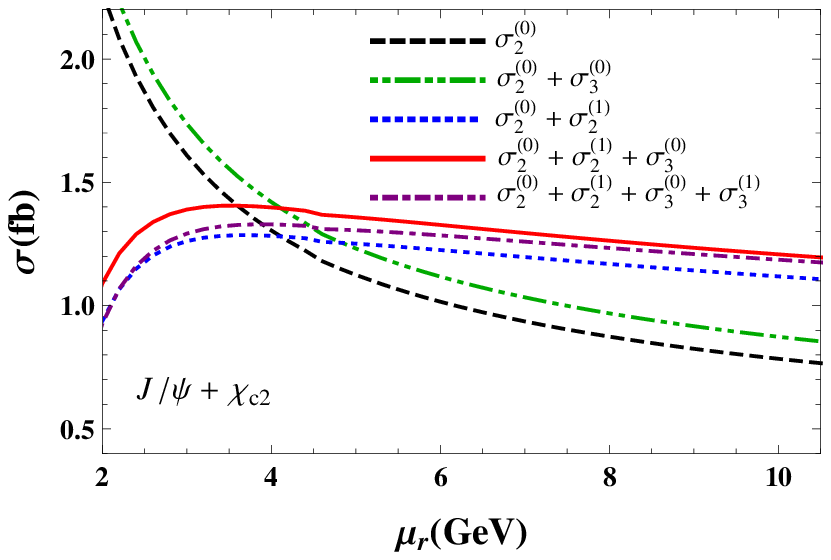}
\caption{The renormalization scale dependence of the total cross sections for $e^+e^- \to J/\psi+\chi_{cJ}$ ($J=0,1,2$) at $\sqrt{s}=10.6$ GeV. $m_c=1.5$ GeV.}
\label{fig:rmu}
\end{figure*}

The $\mu_r$ dependence of the total cross sections for $e^+e^- \to J/\psi+\chi_{cJ}$ ($J=0,1,2$) are illustrated in Figure. \ref{fig:rmu}. As shown in this figure, for the $\chi_{c1}$ and $\chi_{c2}$ cases, the $\mathcal O (\alpha^3\alpha_s^2)$-order terms, $\sigma^{(1)}_3$, can largely counteract the $\sigma^{(0)}_3$ contributions, especially when $\mu_r$ is around 3 GeV, consequently leading to a significant effect on the cross sections of $\sigma^{(0)}_2+\sigma^{(1)}_2+\sigma^{(0)}_3$.

\begin{figure}[htb]
\includegraphics[width=0.59\textwidth]{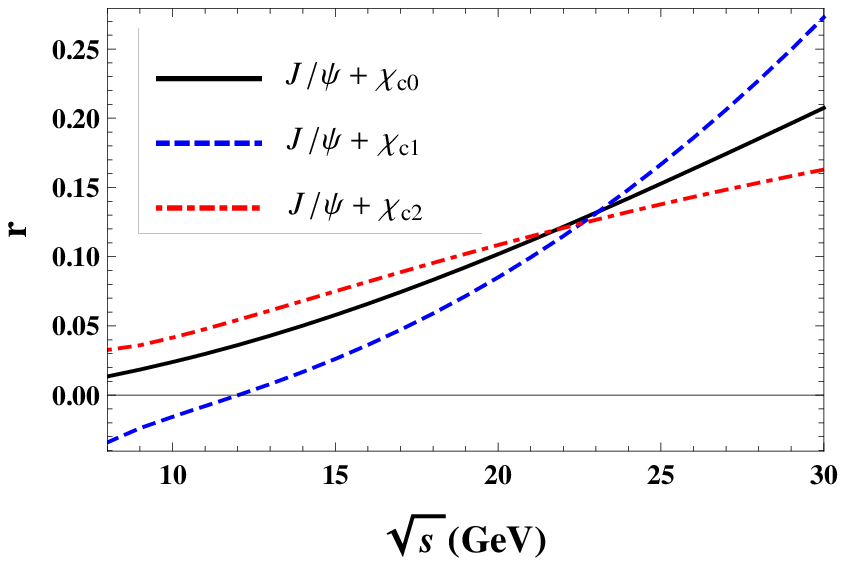}
\caption{$r=\sigma_3/\sigma_2$ as a function of $\sqrt{s}$. }
\label{fig:r}
\end{figure}

To investigate the relative importance of the newly introduced interference terms at higher collision energy, we define the ratio of $r=\sigma_3/\sigma_2$, namely $(\sigma^{(0)}_3+\sigma^{(1)}_3)/(\sigma^{(0)}_2+\sigma^{(1)}_2)$, as a function of $\sqrt{s}$, which is illustrated in Figure \ref{fig:r}, with $\mu_r=\sqrt{s}/2$. As demonstrated in this figure, the ingredient of $\sigma^{(0)}_3+\sigma^{(1)}_3$ will play an more and more important role as the center-of-mass energy rises, especially for the $\chi_{c1}$ case. To be specific, when $\sqrt{s}=30$ GeV, the values of $r$ can reach up to $15\%$, $20\%$ and $26\%$, corresponding to $\chi_{c0}$, $\chi_{c1}$ and $\chi_{c2}$, respectively. Therefore, at the future $e^+e^-$ collider with much higher collision energy, such as the ILC (International Linear Collider) and the $\textrm{Super}-Z$ factory, for the exclusive production of $e^+e^- \to J/\psi+\chi_c$, the interference effect may be fundamental, or even dominant in comparison with the usual QCD contributions.

\section{Summary}

In this paper, by introducing the cross terms between the QCD and QED diagrams, we carry out a further study on the exclusive productions of $e^+e^- \to J/\psi+\chi_{cJ}$ ($J=0,1,2$) at the $B$ factories, based on the NRQCD framework, providing a complete comparison between the interference effects and the usual QCD contributions, at the QCD NLO level, for the first time. It is found that the $\mathcal O (\alpha^3\alpha_s)$-order terms representing the interference effect between the born-level QCD and QED diagrams can provide nonnegligible contributions, which are comparable with the usual NLO QCD corrections, especially for the $\chi_{c1}$ and $\chi_{c2}$ cases. By calculating the QCD corrections to these $\mathcal O (\alpha^3\alpha_s)$-order terms for the first time, we find that the higher order terms, namely $\mathcal O (\alpha^3\alpha_s^2)$, will lead to a significant effect on the $\mathcal O (\alpha^3\alpha_s)$ results. Especially, in the case of $\chi_{c1}$ and $\chi_{c2}$, the newly calculated $\mathcal O (\alpha^3\alpha_s^2)$-order terms can largely counteract the $\mathcal O (\alpha^3\alpha_s)$ contributions. Therefore, to achieve a sound estimate on the total cross sections for $e^+e^- \to J/\psi+\chi_{cJ}$, it is indispensable to include the new $\mathcal O (\alpha^3\alpha_s^2)$-order ingredient. In addition, it is found that, as the collision energy rises, the ratio taken by the interference effect between the QCD and the QED diagrams to the usual QCD cross section will increase rapidly, especially for the $\chi_{c1}$ case.

\hspace{1cm}

\section{Acknowledgments}
\noindent{\bf Acknowledgments}:
This work is supported in part by the Natural Science Foundation of China under the Grant No.11705034., by the Project for Young Talents Growth of Guizhou Provincial Department of Education under Grant No.KY[2017]135, and the Key Project for Innovation Research Groups of Guizhou Provincial Department of Education under Grant No.KY[2016]028.\\

\end{document}